\documentclass[preprint,showpacs,amsmath,amssymb,aps,prd]{revtex4}

\usepackage{graphicx}

\usepackage{amsmath}

 \newcommand{\ds}{\displaystyle}

\begin{document}

\title{A class of regular bouncing cosmologies}

\author{Milovan Vasili\'c} \email{mvasilic@ipb.ac.rs}
\affiliation{Institute of Physics, University of Belgrade, P.O.Box 57,
11001 Belgrade, Serbia}

\date{\today}

\begin{abstract}
In this paper, I construct a class of everywhere regular geometric
sigma models that possess bouncing solutions. Precisely, I show that
every bouncing metric can be made a solution of such a model. My
previous attempt to do so by employing one scalar field has failed due
to the appearance of harmful singularities near the bounce. In this
work, I use four scalar fields to construct a class of geometric sigma
models which are free of singularities. The models within the class
are parametrized by their background geometries. I prove that,
whatever background is chosen, the dynamics of its small perturbations
is classically stable on the whole time axes. Contrary to what one
expects from the structure of the initial Lagrangian, the physics of
background fluctuations is found to carry 2 tensor, 2 vector and 2
scalar degrees of freedom. The graviton mass, that naturally appears
in these models, is shown to be several orders of magnitude smaller
than its experimental bound. I provide three simple examples to
demonstrate how this is done in practice. In particular, I show that
graviton mass can be made arbitrarily small.
\end{abstract}

\pacs{04.50.Kd, 98.80.Jk}

\maketitle

\section{Introduction}\label{Sec1}

The motivation for this work comes from the need for systematization
of the unpleasantly extensive number of cosmological models found in
literature. Indeed, encouraged by the latest astronomical
observations \cite{1, 2, 3, 4, 5, 6, 7, 8, 9, 10, 10.1, 10.2}, the
physical community made a serious effort to model the newly
discovered accelerated expansion of the Universe. As a consequence, a
variety of dark energy models appeared in literature \cite{10.3,
10.4, 10.5, 10.6, 11, 12, 13, 14, 21, 22, 23, 24, 25, 26, 27, 28, 29,
30, 31}. In most of them, the authors search for the inflaton
potential that makes the desired (observationally acceptable)
background metric a solution to their equations.

In this paper, I demonstrate how a freely chosen bouncing metric can
be made a stable solution of a simple geometric sigma model. Geometric
sigma models differ from ordinary sigma models in two respects. First,
all scalar fields can be gauged away, leaving us with a purely metric
theory. Second, its construction follows a rather peculiar way. One
first chooses the metric one would like to be the vacuum of the model,
and then builds a theory that has this metric as its solution. These
models have first been proposed in \cite{32} in the context of
fermionic excitations of flat geometry. Here, I use them for modeling
dark energy dynamics of the Universe. In my previous paper \cite{33},
I considered the simplest geometric sigma models with one scalar
field. Such geometric sigma models successfully generated various
inflationary cosmologies, but their bouncing solutions were mostly
unstable. In this paper, I shall consider geometric sigma models with
$4$ scalar fields. As will shortly be clear, such models are
compatible with the existence of regular and stable bouncing
solutions.

The results obtained in this paper are summarized as follows. First,
a class of purely geometric dark energy models has been constructed.
Every particular model is defined as a geometric sigma model of $4$
scalar fields coupled to gravity. By construction, their background
metrics are spatially flat, homogeneous and isotropic, while scalar
fields are pure gauge. In particular, an arbitrarily chosen bouncing
metric is made a solution of the properly defined geometric sigma
model. Ultimately, one is provided with the class of dark energy
models parametrized by their background geometries.

The second result concerns the absence of singularities in this class
of models. It has been shown that the invertibility of the sigma
model target metric is sufficient to ensure the absence of physical
singularities. In particular, it has been demonstrated that small
perturbations of the background regularly propagate through the
bounce. This holds true irrespectively of the fact that the chosen
sigma model target metric is not positive definite.

The third result establishes linear stability of the background
fluctuations in the considered geometric sigma models. This is the
main result of the paper. It has been proven true for all
background geometries. In particular, whatever bouncing metric is
chosen to be the background, there is a class of scalar field
potentials that makes it classically stable on the whole time axes.
This is a great achievement, as most of the bouncing models found in
literature suffer from the appearance of instability at some moment in
the history of the Universe. (For other bouncing scenarios, see the
recent bouncing literature \cite{33.1, 33.2, 33.3, 33.4, 33.5, 33.6,
33.7, 33.8}, and references therein.)

The fourth result concerns the particle spectrum of the considered
geometric sigma models. As the initial action governs the dynamics of
$4$ scalar fields coupled to Einstein gravity, one expects to have one
graviton and $4$ scalar particles. Surprisingly, what one finds is the
particle spectrum that consists of $2$ tensor, $2$ vector and $2$
scalar degrees of freedom. The graviton mass, that appears as a
byproduct, is shown to be several orders of magnitude smaller than its
experimental bound.

Finally, I have repeated the analysis of Ref. \cite{33} to show that
inclusion of ordinary matter does not compromise the established
background stability. In particular, the presence of a perfect fluid
is shown to modify the background dynamics very much the same as
$\Lambda$CDM model does.

The layout of the paper is as follows. In Sec. \ref{Sec2}, the
construction of geometric sigma models, as defined in \cite{32}, is
recapitulated and subsequently applied to spatially flat, homogeneous
and isotropic geometries. As a result, a class of action functionals
of the Universe is obtained. Each of these action functionals has a
nontrivial background solution that stands for the background
geometry of the Universe. In Sec. \ref{Sec3}, the dynamics of small
perturbations of these nontrivial backgrounds is examined. Despite
the fact that only scalar fields are coupled to the metric in the
initial action, the particle spectrum is found to consist of $2$
tensor, $2$ vector and $2$ scalar degrees of freedom. In Sec.
\ref{Sec3.5}, the apparent singularities found in coefficients of the
linearized field equations are shown to be unphysical. In Sec.
\ref{Sec4}, the background solutions are proven stable for all
spatially flat, homogeneous and isotropic geometries. In Sec.
\ref{Sec5}, the examples of three bouncing Universes are used to
demonstrate how geometric sigma models are constructed in practice.
Sec. \ref{Sec6} is devoted to concluding remarks.

My conventions are as follows. The indexes $\mu$, $\nu$, ... and $i$,
$j$, ... from the middle of the alphabet take values $0,1,2,3$. The
indexes $\alpha$, $\beta$, ... and $a$, $b$, ... from the beginning
of the alphabet take values $1,2,3$. The spacetime coordinates are
denoted by $x^{\mu}$, the ordinary differentiation uses comma
($X_{,\,\mu} \equiv \partial_{\mu} X$), and the covariant
differentiation uses semicolon ($X_{;\,\mu}\equiv \nabla_{\mu}X$).
The repeated indexes denote summation: $X_{\alpha\alpha} \equiv
X_{11}+X_{25}+X_{33}$. The signature of the $4$-metric $g_{\mu\nu}$
is $(-,+,+,+)$, and the curvature tensor is defined as
$R^{\mu}{}_{\nu\lambda\rho} \equiv \partial_{\lambda}
\Gamma^{\mu}{}_{\nu\rho} - \partial_{\rho}
\Gamma^{\mu}{}_{\nu\lambda} + \Gamma^{\mu}{}_{\sigma\lambda}
\Gamma^{\sigma}{}_{\nu\rho} - \Gamma^{\mu}{}_{\sigma\rho}
\Gamma^{\sigma}{}_{\nu\lambda}$.

\section{Geometric sigma models}\label{Sec2}

\subsection{General considerations}\label{Sub2a}

Geometric sigma models are theories constructed out of the predefined
spacetime metric $g_{\mu\nu}^{(o)}(x)$. The metric $g_{\mu\nu}^{(o)}$
is freely chosen, and the coordinates $x^{\mu}$ are fully fixed. This
way, the functional dependence on $x$ in $g_{\mu\nu}^{(o)}(x)$, and
the corresponding Ricci tensor $R_{\mu\nu}^{(o)}(x)$, is completely
determined. I postulate the following Einstein like equations:
\begin{equation} \label{1}
R_{\mu\nu} = R_{\mu\nu}^{(o)}(x) \,.
\end{equation}
The metric $g_{\mu\nu}^{(o)}$ is a solution of the equation
(\ref{1}). In what follows, I shall call it vacuum.

The equation (\ref{1}) obviously lacks general covariance. To
covariantize it, I introduce a new set of coordinates
$\phi^i=\phi^i(x)$. In terms of these new coordinates, the equation
(\ref{1}) takes the form
\begin{equation} \label{2}
R_{\mu\nu} = H_{ij}(\phi) \phi^i_{,\mu} \phi^j_{,\nu} \,,
\end{equation}
where the functions $H_{ij}(\phi)$ are defined through
\begin{equation} \label{3}
H_{ij}(\phi) \equiv R_{ij}^{(o)}(\phi) \,.
\end{equation}
In other words, the ten functions $H_{ij}(\phi)$ are obtained by
replacing $x$ with $\phi$ in ten components of the Ricci tensor
$R_{\mu\nu}^{(o)}(x)$. The equation (\ref{2}) is generally covariant
once the new coordinates $\phi^i$ are seen as scalar functions of the
old coordinates $x^{\mu}$. If the new coordinates are chosen to
coincide with the old ones, $\phi^i(x)\equiv x^i$, the covariant
equation (\ref{2}) is brought back to its non-covariant form
(\ref{1}).

The equation (\ref{2}) has the form of the Einstein's equation in
which four scalar fields $\phi^i(x)$ of some nonlinear sigma model are
coupled to gravity. The "matter field equations" are obtained from the
Bianchi identities $2R^{\mu\nu}{}_{;\nu} = g^{\mu\nu} R_{,\nu}$.
If the condition $\det \phi^i_{,\mu} \neq 0$ is fulfilled, one obtains
\begin{equation} \label{4}
H_{ij}\nabla^2\phi^j = {1\over 2}
\left({{\partial H_{jk}}\over{\partial \phi^i}} -
{{\partial H_{ki}}\over{\partial \phi^j}} -
{{\partial H_{ij}}\over{\partial \phi^k}}\right)
\phi^j_{,\mu}\phi^{k,\mu} .
\end{equation}
The equation (\ref{4}) is not an independent equation, as it follows
from (\ref{2}) and the Bianchi identities. It is straightforward to
verify that the equations (\ref{2}) and (\ref{4}) follow from the
action functional
\begin{equation} \label{5}
I_g = \frac{1}{2\kappa}\int d^4x\sqrt{-g}\left[ R -
H_{ij}(\phi)\phi^i_{,\mu}\phi^{j,\mu} \right] .
\end{equation}
The target metric $H_{ij}(\phi)$ is constructed out of the background
metric $g_{\mu\nu}^{(o)}$, through its defining relation (\ref{3}).
This way, an action functional is associated with every freely chosen
background metric. This action functional describes a nonlinear sigma
model coupled to gravity, and has the nontrivial vacuum solution
\begin{equation} \label{6}
 \phi^i = x^i \,, \quad
g_{\mu\nu} = g_{\mu\nu}^{(o)}\,.
\end{equation}
The physics of small perturbations of the vacuum (\ref{6}) does not
violate the condition $\det \phi^i_{,\mu} \neq 0$, which enables one
to fix the gauge $\phi^i(x)=x^i$. This gauge brings us back to the
non-covariant geometric equation (\ref{1}).

Equation (\ref{1}) is not the unique geometric equation that allows
the solution $g_{\mu\nu} = g_{\mu\nu}^{(o)}$. A simple generalization
of this equation is obtained by adding terms proportional to
$g_{\mu\nu} - g_{\mu\nu}^{(o)}$. The simplest choice is the equation
\begin{equation} \label{7}
R_{\mu\nu} = R_{\mu\nu}^{(o)}(x) + \frac{1}{2} V(x)
\left( g_{\mu\nu} - g_{\mu\nu}^{(o)} \right) .
\end{equation}
It defines a class of geometric theories parametrized by metrics
$g_{\mu\nu}^{(o)}$, and potentials $V$. The covariantization of the
non-covariant equation (\ref{7}) ultimately leads to the action
functional
\begin{equation} \label{8}
I_g = \frac{1}{2\kappa}\int d^4x\sqrt{-g}\left[ R -
F_{ij}(\phi)\phi^i_{,\mu}\phi^{j,\mu} - V(\phi) \right] ,
\end{equation}
where the target metric $F_{ij}(\phi)$ is defined by
\begin{equation} \label{9}
F_{ij}(x)\equiv R_{ij}^{(o)}(x) - \frac{1}{2} V(x)
g_{ij}^{(o)}(x) \,.
\end{equation}
The class of theories defined by (\ref{8}) possesses the vacuum
solution (\ref{6}) for any choice of the potential $V(\phi)$. The
physics of small perturbations of this vacuum allows the gauge
condition $\phi^i = x^i$, which brigs us back to the geometric
equation (\ref{7}).

\subsection{Cosmology}\label{Sub2b}

In what follows, I shall construct a class of geometric sigma models
based on a spatially flat, homogeneous and isotropic metric
$g_{\mu\nu}^{(o)}$, defined by
\begin{equation} \label{10}
ds^2 = -dt^2 + a^2(t)\left( dx^2 + dy^2 + dz^2 \right).
\end{equation}
The nonzero components of the corresponding Ricci tensor take the form
$$
R_{00}^{(o)} = -3\,\frac{\ddot a}{a} \,,   \quad
R_{\alpha\beta}^{(o)} = \left(a\ddot a +
2\dot a^2 \right)\delta_{\alpha\beta} \,,
$$
where ''dot'' denotes time derivative. It is seen that both,
$g_{\mu\nu}^{(o)}$ and $R_{\mu\nu}^{(o)}$, are functions of time
only. If the potential $V(x)$ is also chosen to be independent of
spatial coordinates, so will be the target metric $F_{ij}$. Indeed,
the nonzero components of $F_{ij}$ are found to be
\begin{equation} \label{11}
F_{00} = W - 2\dot H \,, \quad
F_{ab}=-a^2\,W \delta_{ab} \,,
\end{equation}
where $H\equiv \dot a/a$ is the Hubble parameter, and $W$ is defined by
\begin{equation} \label{12}
V \equiv 2\left( W + \dot H + 3H^2 \right) .
\end{equation}
The target metric $F_{ij}(\phi)$, and the potential $V(\phi)$ are
obtained by the substitution $x^i \to \phi^i$ in $F_{ij}(x)$ and
$V(x)$. Owing to their independence of spatial coordinates, this leads
to the target metric and the potential that depend on $\phi^0$ only.
The corresponding action is that of (\ref{8}), with $F_{ij}$ and $V$
defined by (\ref{11}) and (\ref{12}). It governs the dynamics of
gravity coupled to $4$ scalar fields, and has the vacuum solution
(\ref{6}). The precise form of the target metric $F_{ij}(\phi^0)$, and
the potential $V(\phi^0)$ is determined once the functions $a(t)$ and
$W(t)$ are specified. The class of action functionals (\ref{8})
represents a collection of dark energy models parametrized by $a(t)$
and $W(t)$. 

Before I go on, let me mention that the authors of Refs. \cite{33.92,
33.93, 33.94} developed another way to parametrize cosmologies by
scale factors. Their procedure was named ''cosmological
reconstruction'', and was primarily intended for the construction of
$f(R)$ cosmological models. In particular, it was shown that any
bouncing metric could be obtained by a proper choice of the function
$f(R)$. However, the problem with $f(R)$ bounces is that they are
typically unstable. Indeed, it is well known that every $f(R)$ theory
can be rewritten as a scalar-tensor theory with a single scalar field
\cite{33.92, 33.93, 33.94}. These single-scalar theories are known to
be plagued with instabilities for most of their bouncing solutions. In
fact, this is exactly what I demonstrated in my previous paper
\cite{33}, and what lead me to consider multiple scalars in this
paper. It will become clear later that the multi-scalar concept
developed in this paper leads to an everywhere stable dynamics,
irrespectively of the type of bounce considered. Shortly, my class of
geometric sigma models allows {\it any} regular bounce to be the
background solution whose small perturbations have stable propagation
on the whole time axes. This is certainly not the case with F(R)
bounces.

Standard physical requirements that ensure the absence of ghosts and
tachyons restrain the target metric $F_{ij}(\phi)$ to be positively
definite, and the potential $V(\phi)$ to be bounded from below.
Unfortunately, this is not the case with our class of sigma models.
Indeed, the potential $W$ must be everywhere negative to ensure
$F_{ab}>0$. But then, the Hubble parameter $H$ must monotonously
decrease to prevent $F_{00}$ from becoming negative. As a result, the
bouncing solutions are excluded. In what follows, I shall consider
everywhere negative $W$. Aware of the fact that $F_{00}$ cannot be
everywhere positive, I shall demand that $F_{00}$ is everywhere
negative. This way, the target metric $F_{ij}$ becomes an everywhere
invertible matrix ($\det F_{ij} \neq 0$). The conditions
\begin{equation} \label{13}
W<0 \,, \quad
F_{00} < 0
\end{equation}
will be justified later when I demonstrate their necessity for proving
regularity and stability. (The possibility that scalar fields with the
wrong sign in the kinetic term may be observationally allowed has been
considered before \cite{34, 35, 36, 37, 38, 39}.)

In the next section, I shall examine the dynamics of small
perturbations of the vacuum (\ref{6}). It will be shown that the class
of geometric sigma models (\ref{8}) supports regular and stable
bouncing backgrounds, irrespectively of the violation of the standard
physical requirements.

\section{Dynamics of small perturbations}\label{Sec3}

\subsection{Preliminaries}\label{Sub3a}

In this section, I shall examine dynamics of small perturbations of
the vacuum (\ref{6}), as governed by the action functional (\ref{8}).
The variables of the theory are the scalar perturbation $\varphi^i$,
and the metric perturbation $h_{\mu\nu}$. They are defined by
\begin{equation} \label{14}
\phi^i = x^i + \varphi^i \,, \quad
g_{\mu\nu} = g^{(o)}_{\mu\nu} + h_{\mu\nu}\,.
\end{equation}
The infinitesimal change of coordinates $x^{\mu} \to x^{\mu} +
\xi^{\mu}(x)$ leaves the action invariant, and allows for a gauge
fixing. In the gauge $\phi^i = x^i$, the matter field equations are
identically satisfied, and we are left with the gravitational field
equation (\ref{7}). In this analysis, however, I shall use the gauge
condition
\begin{equation} \label{15}
\phi^0 = t\,, \quad
g_{0\alpha} = 0 \,.
\end{equation}
Then, the residual diffeomorphisms are defined by the constraints
\begin{equation} \label{16}
\xi^0 = 0 \,, \quad \dot\xi^{\alpha} = 0 \,.
\end{equation}
With respect to the residual diffeomorphisms, the variables of the
gauge fixed theory transform as
\begin{equation} \label{17}
\begin{array}{lcl}
\ds \delta_0\varphi^{\alpha} &=&  
\ds -\xi^{\alpha} + {\cal O}_2 \,,                         \\ 
\ds \delta_0 h_{\alpha\beta} &=& 
\ds -a^2\left( \xi_{\alpha,\beta} +
\xi_{\beta,\alpha}\right) + {\cal O}_2 \,,              \\ 
\ds \delta_0 h_{00} &=& 
\ds {\cal O}_2 \,,
\end{array}
\end{equation}
where $\delta_0$ is the form variation, and ${\cal O}_2$ denotes
higher order terms. Here, and in what follows, I adopt the convention
to lower spatial indexes by the Kronecker delta. Thus,
$$
\xi_{\alpha}\equiv\delta_{\alpha\beta}\,\xi^{\beta} , \quad
\varphi_a\equiv\delta_{ab}\,\varphi^b , \quad \dots
$$
The field equations obtained by varying the action (\ref{8}) read
\begin{equation}\label{18}
\begin{array}{rrc}
&&\ds R_{\mu\nu}-\frac 12 V g_{\mu\nu} =
F_{ij}\phi^i_{,\mu}\phi^j_{,\nu} \,,                     \\ 
&&\ds 2F_{ki}\,\Box\phi^i + \left( 2F_{ki,j} -
F_{ij,k} \right)\phi^i_{,\mu}\phi^{j,\mu} -V_{,k} = 0 \,. 
\end{array}
\end{equation}
They govern the dynamics of the metric $g_{\mu\nu}$, and the scalar
fields $\phi^i$. To examine the dynamics of their small
perturbations, it suffices to consider only linear terms.

The linearized field equations are obtained by rewriting (\ref{18})
in terms of $h_{\alpha\beta}$ and $\varphi_{\alpha}$, and
subsequently neglecting higher order terms. The straightforward
calculation leads to quite cumbersome expressions, which I choose not
to display here. Instead, I shall first simplify them by decomposing
$h_{\alpha\beta}$ and $\varphi_{\alpha}$ to their irreducible
components with respect to the rotational group. In the first step,
divergences are subtracted in the decomposition
\begin{subequations}\label{19}
\begin{equation}\label{19a}
\begin{array}{rl}
\ds h_{\alpha\beta} = & 
\ds \tilde h_{\alpha\beta} + \tilde h_{\alpha,\beta} +
\tilde h_{\beta,\alpha} + h_{,\alpha\beta} \,,               \\ 
\ds \varphi_{\alpha} = & 
\ds \tilde \varphi_{\alpha} + \varphi_{,\alpha} \,.
\end{array}
\end{equation}
Here, the new variables $\tilde h_{\alpha\beta}$, $\tilde h_{\alpha}$
and $\tilde\varphi_{\alpha}$ are constrained by
$$
\tilde h_{\alpha\beta,\beta}=\tilde h_{\alpha,\alpha} =
\tilde\varphi_{\alpha,\alpha}=0 \,.
$$
(The variable $\varphi_{\alpha}$ is
treated as a vector despite the scalar nature of the original
$\phi^i$. This will be clarified at the end of this
section.) In the second step, the trace is subtracted from $\tilde
h_{\alpha\beta}$. The traceless part of $\tilde h_{\alpha\beta}$ is
defined by
\begin{equation}\label{19b}
\hat h_{\alpha\beta} \equiv \tilde h_{\alpha\beta}
-\frac{1}{2}\,\tilde h_{\gamma\gamma}\delta_{\alpha\beta} +
\frac{1}{2}\partial_{\alpha}\partial_{\beta}\left(\Delta^{-1}
\tilde h_{\gamma\gamma}\right) ,
\end{equation}
\end{subequations}
where $\Delta^{-1}$ stands for the inverse of the Laplacian $\Delta
\equiv \delta^{\alpha\beta}\partial_{\alpha}\partial_{\beta}$. In
what follows, I shall simplify the analysis by the assumption that
metric perturbations are {\it spatially localized}. This means that
the perturbations $h_{\mu\nu}$ and $\varphi^i$ are assumed to
decrease sufficiently fast in spatial infinity. With this assumption,
Laplacian $\Delta$ becomes an invertible operator, and the equations
like $\partial_{\alpha}X=0$, or $\Delta X=0$ have the unique solution
$X=0$. The new variable $\hat h_{\alpha\beta}$ is constrained by
$$
\hat h_{\alpha\alpha} = \hat h_{\alpha\beta,\beta} = 0 \,.
$$

\subsection{Field equations}\label{Sub3b}

With these preliminaries, the linearized field equations are given as
follows. First, the equations
\begin{subequations}\label{20}
\begin{equation}\label{20a}
\left[\frac{1}{a^2}\tilde h_{\alpha\alpha}\right]_{,0} -
2W a^2 \dot\varphi  + 2H h_{00} =0 \,,
\end{equation}
\begin{equation}\label{20b}
\left[\frac{1}{a^2}\tilde h_{\alpha}\right]_{,00} +
3H \left[\frac{1}{a^2}\tilde h_{\alpha}\right]_{,0} +
2W \left( \tilde\varphi_{\alpha} - \frac{1}{a^2}
\tilde h_{\alpha} \right) = 0 \,,
\end{equation}
%
\begin{equation}\label{20c}
\frac{1}{2H}\left[\frac{1}{a^2}
\tilde h_{\alpha\alpha}\right]_{,0} +
\left[\frac{1}{a^2}\tilde h_{\alpha\alpha}\right] -
2a^2 \left[ \ddot\varphi + \left( 5H+\frac{\dot W}{W}+
\frac{W}{2H} \right)\dot\varphi \right] +
\Delta \left( 2\varphi - \frac{1}{a^2}h \right) = 0
\end{equation}
%
\end{subequations}
are used to solve for $h_{00}$, $\tilde\varphi_{\alpha}$  and $h$.
Thus, these variables carry no degrees of freedom. The remaining six
variables are dynamical. They consist of two tensor modes $\hat
h_{\alpha\beta}$, two vector modes $\tilde h_{\alpha}$, and two
scalars $\tilde h_{\alpha\alpha}$ and $\varphi$. Their equations are
as follows.

The tensor modes satisfy the equation
%
\begin{equation}\label{21}
\left[\frac{1}{a^2}\hat h_{\alpha\beta}\right]_{,00} +
3H\left[\frac{1}{a^2}\hat h_{\alpha\beta}\right]_{,0} -
\frac{1}{a^2}\Delta \left[\frac{1}{a^2}
\hat h_{\alpha\beta}\right] - 2W \left[\frac{1}{a^2} 
\hat h_{\alpha\beta}\right] = 0 \,.
\end{equation}
%
Being subject to the constraints $\hat h_{\alpha\alpha} = \hat
h_{\alpha\beta,\beta} = 0$, the variable $\hat h_{\alpha\beta}$
carries two physical degrees of freedom. Note, however, the presence
of the mass term in (\ref{21}). It will be demonstrated later that the
corresponding graviton mass $m_g = \sqrt{-2W}$ can be made ten orders
of magnitude smaller than its experimental bound. In some cases, it is
possible to make it arbitrarily small.

The vector modes are governed by the equation
\begin{equation}\label{22}
\ddot{\tilde\chi}_{\alpha} -
\left(3H+\frac{\dot W}{W}\right)\dot{\tilde\chi}_{\alpha} -
\frac{1}{a^2}\Delta\,\tilde\chi_{\alpha} -
2W \tilde\chi_{\alpha} = 0 \,,
\end{equation}
where ${\tilde\chi}_{\alpha}$ stands for
\begin{equation}\label{23}
\tilde\chi_{\alpha} \equiv a^3
\left[\frac{1}{a^2}\tilde h_{\alpha}\right]_{,0} .
\end{equation}
Although the equation (\ref{22}) is a second order differential
equation with respect to $\tilde\chi_{\alpha}$ it is of third order
with respect to the original variable $\tilde h_{\alpha}$. It seems as
if $\tilde h_{\alpha}$ carried more than two degrees of freedom. This
is, however, not the case. Indeed, after the gauge fixing (\ref{15}),
we are left with the residual gauge symmetry which can further be
fixed. (The residual parameters $\xi_{\alpha}$ are arbitrary functions
of spatial coordinates, alone). It is straightforward to verify that
$\tilde h_{\alpha}/a^2$ transforms as
\begin{equation}\label{24}
\delta_0 \left[ \frac{1}{a^2}\tilde h_{\alpha} \right] =
-\tilde\xi_{\alpha} \,,
\end{equation}
where $\tilde\xi_{\alpha}$ is the divergence free part in the
decomposition
$$
\xi_{\alpha} = \tilde\xi_{\alpha} + \xi_{,\alpha} \,.
$$
The restriction $\dot\xi_{\alpha}=0$ then implies that
$\tilde\chi_{\alpha}$ is gauge invariant. Let me now integrate
(\ref{23}). One finds
$$
\frac{1}{a^2}\tilde h_{\alpha} =
\int^t_0 \frac{\tilde\chi_{\alpha}}{a^3}\,dt + \tilde c_{\alpha} \,,
$$
where $\tilde c_{\alpha}$ is a divergent free, but otherwise arbitrary
function of $\vec x$. The transformation law (\ref{24}) then tells us
that
$$
\delta_0 \tilde c_{\alpha} =
\delta_0 \left[ \frac{1}{a^2}\tilde h_{\alpha} \right] =
-\tilde\xi_{\alpha} \,.
$$
Both, $\tilde c_{\alpha}$ and $\tilde \xi_{\alpha}$, are divergence free
functions of spatial coordinates, only. This enables one to impose the
gauge condition
$$
\tilde c_{\alpha} = 0 \,,
$$
thereby establishing $1-1$ correspondence between
$\tilde\chi_{\alpha}$ and $\tilde h_{\alpha}$. As a consequence, the
equation (\ref{22}) carries exactly two degrees of freedom. After this
gauge fixing, the residual gauge symmetry is defined by
\begin{equation}\label{25}
\xi_{\alpha} = \xi_{,\alpha} \,,\quad \dot\xi = 0 \,.
\end{equation}

The scalar field equations are most complicated. They govern the
dynamics of $\tilde h_{\alpha\alpha}$ and $\varphi$, which I decide
to display in matrix form. First, I define the matrix variable
$$
\Psi \equiv
\left(
\begin{array}{c}
\ds \psi'  \\ 
\ds \psi''
\end{array}
\right) \equiv
\left(
\begin{array}{c}
\ds \frac{1}{a^2}\tilde h_{\alpha\alpha}  \\   [2ex] 
\ds -4 W a^2 \dot\varphi
\end{array}
\right).
$$
Then, the two scalar equations take the matrix form
\begin{equation}\label{26}
\ddot\Psi - \frac{1}{a^2}\Delta\,\Psi +
{\cal F}\,\dot\Psi + {\cal M}\,\Psi = 0 \,,
\end{equation}
where ${\cal F}$ and ${\cal M}$ are matrix valued coefficients.
The friction coefficient ${\cal F}$, and the mass matrix ${\cal M}$ have
the form
\begin{widetext}
$$
{\cal F} \equiv
\left(
\begin{array}{ccc}
\ds 3H - 2\frac{\dot H}{H} +
\frac{\dot F_{00}}{F_{00}} -
\frac{W}{F_{00}}L && \ds -\frac{H}{F_{00}}L        \\  [2.5ex]
\ds 2\frac{W}{F_{00}}\left(2\dot H +
\frac{\ddot H}{H}\right) && \ds
5H + \left(\frac{W}{F_{00}}-1\right)L              \\
\end{array}
\right) ,
$$
$$
{\cal M} \equiv
\left(
\begin{array}{ccc}
\ds -2W\left(1+\frac{2H}{F_{00}}L\right) &&
\ds -\frac{H}{F_{00}}K                             \\  [2.5ex]
\ds 4W\frac{W}{F_{00}}L     &&
\ds 3H\left(\frac{\dot H}{H}-\frac{\dot W}{W}\right)-
\frac{W}{H}\left(3H + \frac{\dot H}{H}\right) +
\frac{W}{F_{00}}K                                   \\
\end{array}
\right),
$$
\end{widetext}
where the shorthand notation
$$
\begin{array}{rl}
&\ds L \equiv 2H + \frac{\dot W}{W} \,,     \\ 
&\ds K \equiv H\bigg[3L - \left(\frac{1}{H}\right)_{,00}\bigg] +
W \bigg[1-\left(\frac{1}{H}\right)_{,0}\bigg]
\end{array}
$$
is introduced for convenience. The field equations (\ref{26}) carry
two physical degrees of freedom, despite the fact that they are third
order differential equations with respect to the original variables.
To verify this, I shall make use of the residual gauge symmetry
(\ref{25}) to show that all but two scalar degrees of freedom are
nonphysical. First, I integrate $\dot\varphi$ to obtain
$$
\varphi = \int^t_0 \dot\varphi dt + c \,,
$$
where $c$ is arbitrary function of spatial coordinates, only. Then,
the transformation law $\delta_0 \varphi = - \xi$ implies that
$\dot\varphi$ is gauge invariant, and consequently, $\delta_0 c = -\xi
$. This allows the final gauge fixing
$$
c=0
$$
which leaves us with {\it no free gauge parameters}. The gauge fixing
$c=0$ establishes $1-1$ correspondence between $\varphi$ and
$\dot\varphi$, and consequently, between $\Psi$ and $\{\varphi,
\tilde h_{\alpha\alpha}\}$. Thus, the equation (\ref{26}) has two
degrees of freedom.

\subsection{Particle spectrum}\label{Sub3c}

The theory considered in this paper is defined by the action
(\ref{8}), which governs the dynamics of four scalar fields coupled to
gravity. One would expect the particle content of the theory to be
{\it 2 tensor + 4 scalar} degrees of freedom. However, the preceding
results suggest the structure {\it 2 tensor + 2 vector + 2 scalar}
physical fields. Let me clarify this situation. The nature of physical
fields is determined by their transformation properties with respect
to symmetries of the vacuum. Usually, it is the $SO(3,1)$ invariant
Minkowski vacuum that leads to the well known classification of
elementary particles. In cosmology, however, the vacuum is $SO(3)$
invariant. Indeed, the cosmological background is only spatially
homogeneous and isotropic, while its time dependence is nontrivial.
Thus, in what follows, I shall consider the rotational group, defined by
\begin{equation}\label{27}
\xi^{\alpha} = \omega^{\alpha}{}_{\beta}\, x^{\beta} \,.
\end{equation}
The parameters $\omega_{\alpha\beta} = - \omega_{\beta\alpha}$ are
constant and antisymmetric. With respect to diffeomorphisms $x^{\mu}
\to x^{\mu}+\xi^{\mu}$, the scalars $\phi^i$ and the metric
$g_{\mu\nu}$ transform as
$$
{\arraycolsep 0.2em
\begin{array}{rcl}
\ds \delta_0 \phi^i &=& \ds -\xi^{\rho} \phi^i_{\,,\rho}\,,  \\
\ds \delta_0 g_{\mu\nu} &=& \ds
-\xi^{\rho}_{\,,\mu}\, g_{\rho\nu} -
\xi^{\rho}_{\,,\nu}\, g_{\rho\mu} -
\xi^{\rho} g_{\mu\nu,\rho} \,. 
\end{array}}
$$
Their small perturbations, however, transform differently. Indeed, in
the gauge (\ref{15}), their transformation law with respect to
rotations (\ref{27}) becomes
$$
{\arraycolsep 0.2em
\begin{array}{rcl}
\ds \delta_0 \varphi^{\alpha} &=& \ds -\xi^{\alpha} -
\xi^{\beta} \varphi^{\alpha}_{\,,\beta}\,,                           \\
\ds \delta_0 h_{00} &=& \ds -\xi^{\alpha}h_{00,\alpha}\,, \\
\ds \delta_0 h_{\alpha\beta} &=& \ds
-\xi^{\gamma}_{\,,\alpha}\, h_{\gamma\beta} -
\xi^{\gamma}_{\,,\beta}\, h_{\gamma\alpha} -
\xi^{\gamma} h_{\alpha\beta,\gamma} \,.
\end{array}}
$$
It is seen that $h_{\alpha\beta}$ transforms as a tensor, while
$h_{00}$ and $\varphi^{\alpha}_{\,,\alpha}$ are scalars. The variable
$\varphi^{\alpha}$, on the other hand, is neither a vector nor a
collection of three scalars. This state of affairs can change once we
realize that the action (\ref{8}) has an {\it extra global symmetry}.
Indeed, it is easily shown that the action (\ref{8}) is invariant with
respect to
$$
\delta_1 \phi^a = \epsilon^a{}_b\, \phi^b , \quad
\delta_1 \phi^0 = \delta_1 g_{\mu\nu} = 0 \,,
$$
where $\epsilon_{ab}=-\epsilon_{ba}$ are constant parameters
independent of $\omega_{ab}$. The full global symmetry is then
defined by the total variation $\delta_0 + \delta_1$. Let us consider
the subgroup defined by
$$
\epsilon_{ab} = \omega_{ab} \,.
$$
Its action on the variables $\varphi^{\alpha}$ and $h_{\alpha\beta}$
is given by
$$
{\arraycolsep 0.2em
\begin{array}{rl}
\ds \big[\delta_0(\omega)+\delta_1(\omega)\big] \varphi^{\alpha}= &
\ds \xi^{\alpha}_{\,,\beta}\,\varphi^{\beta} -
\xi^{\beta} \varphi^{\alpha}_{\,,\beta} \,,                   \\ 
\ds \big[\delta_0(\omega)+\delta_1(\omega)\big] h_{\alpha\beta}= &
\ds -\xi^{\gamma}_{\,,\alpha}\, h_{\gamma\beta} -
\xi^{\gamma}_{\,,\beta}\, h_{\gamma\alpha} -
\xi^{\gamma} h_{\alpha\beta,\gamma}
\end{array}}
$$
where $\xi^{\alpha} \equiv \omega^{\alpha}{}_{\beta}\, x^{\beta}$. It
is seen that $\varphi^{\alpha}$ transforms as a vector, and
$h_{\alpha\beta}$ as a tensor with respect to global rotations.
Therefore, the terminology {\it "irreducible representations of the
rotational group"}, used for the description of decomposition
(\ref{19}), is justified. As a consequence, the particle
spectrum has the structure
$$
\mbox{\it 2 tensor + 2 vector + 2 scalar}
$$
degrees of freedom, in contrast with what one might expect from the
action (\ref{8}).

\section{Singularities}\label{Sec3.5}

In the preceding section, we have seen that the evolution of small
physical perturbations is governed by the equations (\ref{21}),
(\ref{22}) and (\ref{26}). Let me examine their regularity.

In the first step, one examines if the coefficients of these
equations are everywhere regular. It is straightforward to verify
that the conditions (\ref{13}) ensure the full regularity of the
tensor and vector equations, (\ref{21}) and (\ref{22}). This leaves
us with the scalar equations (\ref{26}), whose coefficients become
singular in $H=0$. This is an improvement with respect to Ref.
\cite{33} because the real harmful singularities $F_{00}=0$ do not
appear in the present approach. In fact, it has been suggested in
\cite{33} that singularity of the bounce ($H=0$) is not physical. In
what follows, I shall demonstrate that the propagation of scalar
modes in the vicinity of the bounce is regular, irrespectively of the
apparent singularity of the $\cal F$ and $\cal M$ coefficients.

Let me first choose the moment of the bounce as a natural origin of
time coordinate. Then, the singular point $H=0$ is identified with
$t=0$. Without loss of generality, I shall consider the scale factor
$a(t)$ that behaves as
$$
a(t) = a_0 + a_2\, t^2 + {\cal O}(t^4)
$$
in the vicinity of $t=0$. (The only reason for dropping term
proportional to $t^3$ is to simplify cumbersome expressions.) A
straightforward calculation then shows that $\cal F$ and $\cal M$
take the form
$$
{\cal F} = -\frac{2}{t}
\left(
\begin{array}{cc}
1 & 0    \\
0 & 0    \\
\end{array}
\right)
+ {\cal O}_0 \,, \ \   
{\cal M} = -\frac{1}{t}
\left(
\begin{array}{cc}
0 & 1    \\
0 & 0    \\
\end{array}
\right)
+ {\cal O}_0 \,,
$$
where ${\cal O}_0$ stands for regular terms. Let me now rewrite the
equation (\ref{26}) in terms of the new variable $\Psi_1$, defined as
\begin{equation}\label{28.5}
\Psi_1 = U^{-1} \Psi =
\left(
\begin{array}{ccc}
\ds 1 && \ds  \frac{t}{2}    \\
\ds 0 && \ds 1
\end{array}
\right) \Psi \,.
\end{equation}
With this, the equation (\ref{26}) takes the form
\begin{equation}\label{29}
\ddot\Psi_1 - \frac{1}{a^2}\Delta\,\Psi_1 +
{\cal F}_1\,\dot\Psi_1 +
{\cal M}_1\,\Psi_1 = 0 \,,
\end{equation}
where ${\cal F}_1$ and ${\cal M}_1$ are given by
\begin{equation}\label{29.5}
\begin{array}{rl}
\ds {\cal F}_1 = & 
\ds U^{-1} \left( {\cal F} U + 2\dot U \right),         \\
\ds {\cal M}_1 = & 
\ds U^{-1}\left({\cal M}U+{\cal F}\dot U+\ddot U\right)
\end{array}
\end{equation}
for every nonsingular $U$. In the case under consideration, one finds
$$
{\cal F}_1 = {\cal F} + {\cal O}_0 \,, \ \ 
{\cal M}_1 = 
\left(\begin{array}{ccc}
\ds -2W && \ds 0                          \\   
\ds 4\dot W \frac{W}{F_{00}} && \ds 3\dot H-2W 
\end{array} \right) + {\cal O}_1 \,.
$$
It is seen that the mass matrix ${\cal M}_1$ is everywhere regular,
so that ${\cal F}_1$ remains the only coefficient with singular
behavior. Now, I am ready to solve the equation (\ref{29}) in the
vicinity of $t=0$. First, I use the Fourier decomposition
$$
\Psi_1 = {\rm Re} \int d^3 k\, Q(k,t)\, e^{i \vec k\cdot\vec x}
$$
to rewrite the equation (\ref{29}) in terms of its Fourier modes:
\begin{equation}\label{31}
\ddot Q + {\cal F}_1\, \dot Q +
\left( {\cal M}_1 + \frac{k^2}{a^2} \right) Q =0 \,.
\end{equation}
In the vicinity of $t=0$, the coefficients ${\cal F}_1$ and ${\cal
M}_1$ are well approximated by
$$
{\cal F}_1 \approx
-\frac{2}{t}\, I \,, \quad
{\cal M}_1 \approx J \,,
$$
where constant matrices $I$ and $J$ are defined by
$$
I \equiv
\left(
\begin{array}{cc}
1 & 0    \\
0 & 0    \\
\end{array} \right) , \quad
J \equiv \left({\cal M}_1 + \frac{k^2}{a^2} \right)_{t=0} \,.
$$
The matrix elements of the matrix valued coefficients $I$ and $J$ are
all finite.  Using this approximation, and multiplying (\ref{31}) by
$t$, one finally arrives at
\begin{equation}\label{32}
t\ddot Q -2 I \dot Q + t J Q =0 \,.
\end{equation}
The solution of the equation (\ref{32}) is searched for in the
form of the power series
$$
Q = \sum_{n=0}^{\infty} q_n t^n \,.
$$
One straightforwardly finds
\begin{subequations}\label{33}
\begin{equation}\label{33a}
I q_1 =0 \,, \quad J q_1 +
\left(
\begin{array}{cc}
0 & 0    \\
0 & 6    \\
\end{array}
\right) q_3 = 0 \,,
\end{equation}
and for $n\neq 1$,
\begin{equation}\label{33b}
q_{n+2} = -\frac{1}{n+2}
\left(
\begin{array}{cc}
\ds\frac{1}{n-1} &             0           \\
0                   & \ds\frac{1}{n+1}    \\
\end{array} \right) J q_n  \,.
\end{equation}
\end{subequations}
The equations (\ref{33b}) tell us that all the coefficients $q_n$ are
determined in terms of $q_0$, $q_1$ and $q_3$. The coefficients $q_0$,
$q_1$ and $q_3$, on the other hand, are constrained by (\ref{33a}),
but are not completely determined. Let us see how many degrees of
freedom we are left with. To this end, I shall use the notation
$$
q_n \equiv
\left(
\begin{array}{c}
q_n'     \\
q_n''
\end{array}
\right)
$$
to rewrite the equations (\ref{33a}) in the component form.
Thus, one finds
\begin{equation}\label{34}
q_1' = 0 \,, \quad
6\,q_3'' + J_{22}\, q_1'' = 0 \,.
\end{equation}
Now, we clearly see that the components $q_0'$, $q_0''$, $q_1''$ and
$q_3'$ remain undetermined. Thus, there exists a class of regular
solutions to the singular equation (\ref{31}), parametrized by four
free parameters
$$
q_0' \,,\quad q_0'' \,,\quad  q_1'' \quad
\mbox{and} \quad  q_3' \,.
$$
These four parameters stand for $2$ physical degrees of freedom. As a
consequence,
\begin{itemize}
\item {\it the class of cosmological models considered in this paper
is everywhere regular}.
\end{itemize}

Let me emphasize that it is very important to have the full number of
degrees of freedom in $t=0$. If it is not the case, only very special
initial conditions in the past lead to trajectories that regularly
pass the bounce. My result is that solutions regular in $t=0$
carry the maximal number of degrees of freedom allowed by the model.
Only then {\it every perturbation formed in the past regularly passes
the bounce}. I have verified this result by numerically solving
differential equations with this kind of singularity. The conditions
(\ref{34}) have also been checked. In particular, I have shown that
$Q'(t)$, formed in the past, always approaches $t=0$ as a constant
($q_1'=0$). Owing to the condition $q_1'=0$, the nonphysical variables
$h_{00}$, $h$ and $\tilde\varphi_{\alpha}$, obtained by solving
equations (\ref{20}), are also everywhere regular.

Finally, I want to draw your attention to the structure of initial
conditions that can be chosen in $t=0$. As opposed to $t\neq 0$, the
initial conditions in $t=0$ cannot be the value of the field and of
its first time derivative. Indeed, while initial conditions for the
scalar $Q''(t)$ can be chosen in the standard way ($q_0''$, $q_1''$),
the scalar $Q'(t)$ is determined by its value $q_0'$, and the value of
its third time derivative $q_3'$.

\section{Stability analysis}\label{Sec4}

\subsection{Tensor and vector modes}\label{Sub4a}

In this section, I shall examine stability of the vacuum against its
small  perturbations, as governed by the equations (\ref{21}),
(\ref{22}) and (\ref{26}). It is immediately seen that both, {\it tensor
and vector equations, have stable dynamics} for any $a(t)$, and any
$W$ that respects the conditions (\ref{13}). Indeed, the two mass
terms are everywhere positive, leading to an oscillatory behavior of
small vacuum perturbations. What one might see as a problem is that
tensor modes, which are usually identified with the graviton, have
nonzero mass
\begin{equation}\label{34.1}
m_g = \sqrt{-2W} \,.
\end{equation}
This is not what one would like to have. Notice, however, that the
term $\sqrt{-2W}$ is time dependent, and therefore, cannot stand for
the conventional mass on the whole time axis. Instead, one should
examine the behavior of $\sqrt{-2W}$ at the present epoch. At the
present epoch, the spacetime is close to being flat, and has rather
slow expansion rate. In a situation like this, the term $\sqrt{-2W}$
can well be considered the graviton mass.

The estimation of the present value of $m_g$ can be done by making
use of the restrictions (\ref{13}). One finds
$$
m^2_g > 4\left( q+1 \right)H^2 \,,
$$
where the Hubble parameter $H$, and deceleration parameter $q$ are
defined by
\begin{equation}\label{34.2}
H \equiv \frac{\dot a}{a} \,, \quad
q \equiv - \frac{\ddot a}{a H^2}  \,.
\end{equation}
The observed values of the Hubble and deceleration
parameters are
\begin{equation}\label{34.3}
H_0 \approx 1.6 \cdot 10^{-33}\, eV , \quad
q_0 \approx - 0.5 \,,
\end{equation}
so that the present graviton mass obeys
$$
m_g > 2.2 \cdot 10^{-33}\, eV \,.
$$
This is more than ten orders of magnitude smaller than the upper
bound reported by the LIGO experiment \cite{45}:
$$
m_g < 1.2 \cdot 10^{-22}\, eV \,.
$$
Thus, there is plenty of room for choosing a plausible cosmological
model from the class of models described in this paper. What one
should do is to make a proper choice of the potential $W$. I shall
demonstrate this in examples of the next section.

Let me mention one more thing about the estimation of graviton mass.
It is seen that $m_g$ is sensitive to the actual value of the
deceleration parameter $q$. In particular, the graviton mass can be
made arbitrarily close to zero if $q<-1$. There is a good reason why
one might believe in such a scenario. Namely, the realistic Universe,
where all the measurements are done, is filled with ordinary matter,
too. The presence of ordinary matter increases the measured value of
$q$. Thus, it may happen that the vacuum value of $q$ is either equal
to $-1$ ($\Lambda$CDM model), or even smaller than $-1$. In the
examples of the next section, I shall demonstrate how the presence of
matter increases the vacuum value $q<-1$ to the measured value
$q=-1/2$.

\subsection{Scalar modes}\label{Sub4b}

The stability of the scalar equation (\ref{26}) is examined by
solving the eigenvalue problem of the mass matrix $\cal M$. What one
hopes to find is that both eigenvalues of $\cal M$ are real and
positive. Only then the vacuum solution of (\ref{26}) is stable
against its small perturbations.

Skipping unnecessary details, the conditions that ensure reality
and positivity of the mass eigenvalues have the form
\begin{equation}\label{35}
\det{\cal M} \geq 0 \,,\quad
{\cal M}_{11} + {\cal M}_{22} - 2\sqrt{\det{\cal M}} \geq 0 \,.
\end{equation}
The conditions (\ref{35}) are not satisfied for every $W$ and $a$.
However, it can be shown that, for every $a(t)$, there exists a class
of potentials $W(t)$ such that (\ref{35}) holds true. As an example,
let me consider $W$ defined by
\begin{equation}\label{36}
W = - \frac{\omega^2}{a^2} \,,
\end{equation}
where $\omega$ is constant with the dimension of mass. Naturally, this
choice of the potential must preserve the regularity conditions
(\ref{13}). If $\omega$ is chosen large enough, the conditions
(\ref{13}) are satisfied for every $a(t)$ with linear or slower growth
in the asymptotic region. For scale factors that grow faster, one may
be led to make a different choice of the potential. However, the
choice (\ref{36}) always respects (\ref{13}) in a finite, arbitrarily
chosen time interval $-T<t<T$. This allows one to keep most of the
potential (\ref{36}), and just modify its asymptotic behavior.

In what follows, I shall use the variable $\Psi_1$, defined in
(\ref{28.5}). This simplifies the form of the mass matrix. Using
(\ref{36}), the mass matrix ${\cal M}_1$ becomes
$$
{\cal M}_1 = \frac{1}{2}
\left(
\begin{array}{cc}
\ds -4W & \ds -3H + 3\,t\left( \dot H + 2H^2 \right)    \\ 
\ds   0 & \ds -4W + 6\left( \dot H + 2H^2 \right)     \\
\end{array}
\right) .
$$
It is now easy to check the conditions (\ref{35}). Indeed, one
straightforwardly finds
$$
\begin{array}{rcl}
\ds \left({\cal M}_1\right)_{11} &=&\ds -2W > 0 \,,    \\
\ds \left({\cal M}_1\right)_{22} &=&\ds \frac{1}{2}
\left( 12H^2 - 3F_{00} - W \right) > 0
\end{array}
$$
whenever the conditions (\ref{13}) are fulfilled. As a consequence,
the determinant is also positive,
$$
\det{\cal M}_1 = \left({\cal M}_1\right)_{11}
\left({\cal M}_1\right)_{22} > 0 \,,
$$
and so is the second of the conditions (\ref{35}):
$$
\begin{array}{rl}
& \ds \left({\cal M}_1\right)_{11} + 
\left({\cal M}_1\right)_{22} - 2\sqrt{\det{\cal M}_1}  \\ 
& {}=\ds \left[ \sqrt{\left({\cal M}_1\right)_{11}} -
\sqrt{\left({\cal M}_1\right)_{22}}\right]^2 > 0 \,.
\end{array}
$$
Thus, we have proved that the {\it mass matrix ${\cal M}_1$ has two
real positive eigenvalues} whenever the potential $W$ has the form
(\ref{36}). This certainly holds true in any finite time interval, and
for any choice of the scale factor $a(t)$. As for the asymptotic
region, it is not difficult to show that $W$ can always be chosen to
ensure the asymptotic stability of the scalar modes. For example, let
us consider the power law behavior of $a(t)$ and $W(t)$. Starting with
$$
a \sim t^{\alpha} , \quad W \sim t^{\beta} ,
$$
one straightforwardly finds that
$$
\left( 2\alpha+\beta \right)\left(\alpha-1\right) > 0
$$
ensures the validity of (\ref{35}). In particular, for all $a(t)$
that grow faster than $t$ in the asymptotic region, it is enough to
choose $W \sim 1/t^2$. The faster the expansion of the Universe is,
the smaller values of $W$ are allowed. In some cases, it is possible
to define $W$ with arbitrarily fast approach to zero. To conclude, I
have proven in this section that
\begin{itemize}
\item {\it for every scale factor $a(t)$, there is a class of
potentials $W(t)$ that makes the dynamics linearly stable}.
\end{itemize}

\subsection{Matter fields}\label{Sub4c}

So far, I have proved regularity and stability of the class of dark
energy models in which ordinary matter has been neglected. One may
wonder if the presence of ordinary matter might spoil the nice
results obtained so far. A similar problem has already been studied
in \cite{33}, with the result that matter fields do not compromise
earlier results. There, only one scalar field has been coupled to
gravity. Nevertheless, the present problem turns out to be
technically identical to that of Ref. \cite{33}. For that reason, I
shall present a short version of the full analysis.

Let me consider the action
\begin{equation}\label{37}
I = I_g + I_m \,,
\end{equation}
where $I_g$ is the geometric action (\ref{8}), and $I_m$ stands for
the action of matter fields. Usually, the matter Lagrangian is taken
to be that the standard model of elementary particles, minimally
coupled to gravity. Matter fields are collectively denoted by
$\Omega$. Owing to the minimal coupling to the metric, the matter
field equations
$$
\frac{\delta I_m}{\delta\Omega} = 0
$$
are trivially satisfied by
$$
\Omega =\Omega_0 \,,
$$
where $\Omega_0$ stands for the well known vacuum of the standard
model of elementary particles. The vacuum value of the stress-energy
tensor $T^{\mu\nu}_m$ is also zero. Formally,
$$
\Omega = \Omega_0 \quad \Rightarrow \quad T^{\mu\nu}_m = 0 \,.
$$
With this, the inflaton and Einstein's equations reduce to those
considered in the preceding sections. Indeed, owing to the absence of
the direct matter-inflaton couplings, the inflaton equations take the
form
$$
\frac{\delta I}{\delta \phi^i} =
\frac{\delta I_g}{\delta \phi^i} = 0 \,,
$$
while Einstein's equations become
$$
R^{\mu\nu}-\frac{1}{2}\,g^{\mu\nu}R = T^{\mu\nu}_{\phi} \,,
$$
whenever $\Omega = \Omega_0$. As a consequence, the model
(\ref{37}) has the vacuum solution
\begin{equation}\label{39}
g_{\mu\nu}=g^{(o)}_{\mu\nu}\,,\quad
\phi^i = x^i \,, \quad
\Omega = \Omega_0 \,.
\end{equation}
Thus, the presence of matter fields does not compromise the
sigma model vacuum of the preceding sections.

The linear stability of the vacuum (\ref{39}) is examined by
inspecting the linearized field equations of the action (\ref{37}).
It is immediately seen that, after linearization, the inflaton and
Einstein's equations reduce to those of the geometric sigma model of
the preceding sections. Indeed, the stress-energy tensor
$T^{\mu\nu}_m$, being at least quadratic in perturbations of matter
fields, does not appear on the r.h.s. of the linearized Einstein's
equations. At the same time, the inflaton does not couple to matter
fields, at all. Hence, the linearized inflaton and metric equations
of motion remain unchanged by the inclusion of matter. They are
diffeomorphism invariant, so that the complete gauge fixing procedure
of Sec. \ref{Sec3} is still valid. In this gauge, the tensor, vector
and scalar equations obtained from $I$ coincide with those obtained
from the geometric action $I_g$. Their linear stability has already
been proven. As for the stability of matter itself, it is enough to
recall that matter field equations reduce to the standard model of
elementary particles, in the inertial reference frame. To summarize,
the vacuum (\ref{39}) is linearly stable against small perturbations
governed by the action (\ref{37}). The dynamics of its geometric part
remains the same as found in Sec. \ref{Sec3}. In conclusion,
\begin{itemize}
\item {\it the presence of matter fields does not violate the
established linear stability}.
\end{itemize}
As a final step, one might consider the inclusion of direct
matter--inflaton couplings. Skipping unnecessary details, I shall
only emphasize that typical matter--inflaton couplings preserve the
regularity and stability arguments given earlier. In particular, the
interaction terms which are at least quadratic in matter fields do
not compromise the previous analysis. More details can be found in
Ref. \cite{33}.

\section{Examples}\label{Sec5}

In this section, I shall analyze three simple models of the Universe.
In the first, a toy bouncing model is used for the demonstration of
how geometric sigma models are constructed in practice. In the
second, I consider a slowly contracting Universe with an exponential
expansion after the bounce. The influence of dust matter is studied
as an example of how ordinary matter modifies the background
dynamics. The third example demonstrates how graviton mass can be
made arbitrarily small.

\subsection{Toy model}\label{Sub5a}

\begin{figure}[htb]
\begin{center}
\includegraphics[height=4cm]{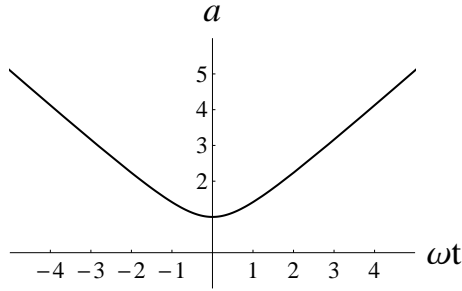}
\end{center}
\vspace*{-.5cm}
\caption{Toy model.\label{f1} }
\end{figure}

Here, I shall study a homogeneous, isotropic and spatially flat
geometry with the scale factor of the form
\begin{subequations}\label{40}
\begin{equation}\label{40a}
a(t) = \sqrt{1+\omega^2 t^2} \,,
\end{equation}
and the potential defined by
\begin{equation}\label{40b}
W(t) = -2\,\frac{\omega^2}{a^2} \,.
\end{equation}
\end{subequations}
Its graph is displayed in Fig. \ref{f1}. It describes a linearly
contracting Universe with the bounce at $t=0$, and linear expansion
afterwards.
The constant $\omega$ is a free parameter with the dimension of mass.
The scale factor (\ref{40a}) is a solution of the sigma model
(\ref{8}) in which the target metric $F_{ij}(\phi)$ and the potential
$V(\phi)$ are defined by the replacement $x^i \to \phi^i$ in the
expressions (\ref{11}) and (\ref{12}). As neither $F_{ij}(x)$ nor
$V(x)$ depend on spatial coordinates, the resulting target metric
\begin{subequations}\label{41}
\begin{equation}\label{41a}
F_{00} = - \frac{4\,\omega^2}{\left(1 +
\omega^2\phi^2_0\right)^2} \,, \quad
F_{ab} = 2\,\omega^2 \delta_{ab} \,,
\end{equation}
and the potential
\begin{equation}\label{41b}
V =  - \frac{2\,\omega^2}{\left(1 +
\omega^2\phi^2_0\right)^2}
\end{equation}
\end{subequations}
depend on $\phi_0$, only. The equations (\ref{40b}) and (\ref{41a})
show that regularity conditions (\ref{13}) are satisfied on the whole
time axis. Then, the general arguments of the preceding sections
ensure that small perturbations of the vacuum have nonsingular, and
everywhere stable dynamics.

The free parameter $\omega$ is determined from the observed values of
the Hubble and deceleration parameters, as given by (\ref{34.3}).
First, we use (\ref{40}) to obtain
$$
\omega = \frac{1-q_0}{\sqrt{-q_0}}\, H_0 \,, \quad
t_0 = \frac{1}{(1-q_0)H_0} \,,
$$
where $t_0$ stands for the present time.
Then, the substitution of (\ref{34.3}) yields
$$
\omega \approx 3.4 \cdot 10^{-33}\ {\rm eV}  \,, \quad
t_0 \approx 8.9 \ {\rm Gyr}  \,.
$$
With these values of $\omega$ and $t_0$, the contemporary value of
the graviton mass $m_g = \sqrt{-2W}$ becomes
$$
m_g \approx 3.6 \cdot 10^{-33}\ {\rm eV} \,,
$$
which is more than ten orders of magnitude smaller than the upper
bound reported by the LIGO experiment \cite{45}. In what follows,
I shall demonstrate how this value of $m_g$ can be made even smaller.

\subsection{Simple bouncing Universe}\label{Sub5b}

\begin{figure}[htb]
\begin{center}
\includegraphics[height=4cm]{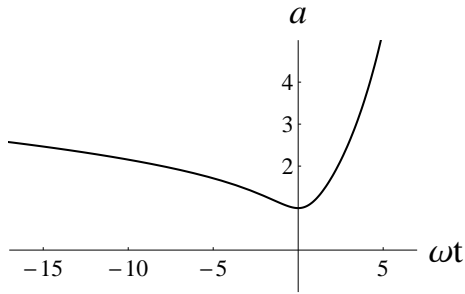}
\end{center}
\vspace*{-.5cm}
\caption{Simple bouncing Universe.\label{f2} }
\end{figure}

In the second example, I shall examine the scale factor
\begin{subequations}\label{42}
\begin{equation}\label{42a}
a(t) = \sqrt[3]{\ds e^{\,\omega t}-\omega t}
\end{equation}
with the potential
\begin{equation}\label{42b}
W = -\frac{\omega^2}{a^2} \,.
\end{equation}
\end{subequations}
The corresponding background dynamics is shown in Fig. \ref{f2}. It
is a slowly contracting Universe with an exponential expansion after
the bounce. In this respect, its late time behavior resembles that of
the $\Lambda$CDM model.
%
%
The target metric and the potential, needed for the construction of
the action (\ref{8}), are obtained by the replacement $x^i \to
\phi^i$ in the expressions (\ref{11}) and (\ref{12}). Skipping the
details of the calculation, I shall only emphasize that $F_{00}$ is
easily checked to be strictly negative. Thus, the reality conditions
(\ref{13}) are everywhere satisfied. As a consequence, the dynamics
of small perturbations of the background (\ref{42a}) is regular and
stable at all times.

\begin{figure}[htb]
\begin{center}
\includegraphics[height=4cm]{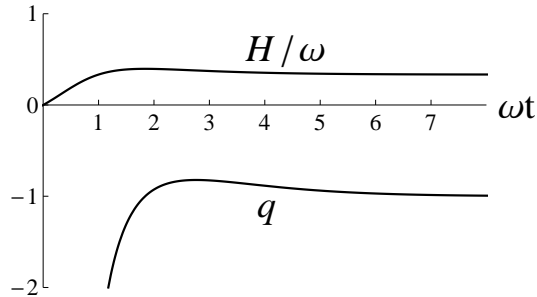}
\end{center}
\vspace*{-.5cm}
\caption{Hubble and deceleration parameters.\label{f3} }
\end{figure}

The parameter $\omega$ is determined from the measured values of the
Hubble and deceleration parameters. To avoid cumbersome expressions,
I shall do this by a graphical method. The graphs of $H(t)$ and
$q(t)$ are displayed in Fig. \ref{f3}.
%
%
As one can see, for positive values of $H$, the deceleration parameter
$q$ can never get close to the measured value $q=-0.5$. There are
three possibilities of how to handle this situation. The first is to
accept the closest allowed value of $q$, which is, in this case,
$q\approx -0.82$. However, this is not well justified, as there is no
reason to have such a confidence in a model that is still under
construction. Instead, one can reformulate the model to allow for the
needed value of $q$. This sounds better, but it has its drawbacks. For one
thing, ordinary matter has to be included. As ordinary matter
increases the value of $q$, it will spoil the chosen vacuum value
$q=-1/2$. In fact, one should start with the smaller vacuum value to
reach the needed $q=-1/2$ after the inclusion of matter. But we
already have that smaller value, namely $q\approx -0.82$. Therefore,
we should check how it is modified by the inclusion of matter.

Let me consider the simple case of a perfect fluid. Its stress-energy
tensor  has the form
$$
T^m_{\mu\nu} =
\left( \rho + p \right)u_{\mu}u_{\nu} + p\,g_{\mu\nu} \,,
$$
where $\rho$ and $p$ stand for the fluid energy density and pressure,
and $u^{\mu}$ is the fluid $4$-velocity. In the case of spatially
homogeneous and isotropic spaces, $u^{\alpha} = 0$ and $u^0 = 1$. In
this example, I shall work in the gauge $\phi^i = x^i$. In this
gauge, the field equations take the form
$$
R_{\mu\nu}-\frac{1}{2}\,g_{\mu\nu}R = \kappa T_{\mu\nu} \,,
$$
where $\kappa \equiv 8\pi G$ is the gravitational constant,
$$
T_{\mu\nu} \equiv T^m_{\mu\nu} +
\frac{1}{\kappa }T^{\phi}_{\mu\nu} \,,
$$
and
$$
T^{\phi}_{\mu\nu} \equiv
F_{\mu\nu} - \frac{1}{2} \left(V + g^{\rho\sigma}
F_{\rho\sigma}\right) g_{\mu\nu} \,.
$$
In the absence of matter, these field equations reduce to the
geometric equations (\ref{7}).

In what follows, I shall make use of the fact that $T_{\mu\nu}$ can
be rewritten as
$$
T_{\mu\nu} =
\left(\bar\rho+\bar p\right)u_{\mu}u_{\nu} + \bar p\, g_{\mu\nu} \,,
$$
where the effective energy density and pressure read
$$
\begin{array}{l}
\ds \bar\rho \equiv \rho + \frac{1}{2\kappa} \left( V + F_{00} -
3W\frac{a^2}{\bar a^2} \right) ,     \\   [1.4ex]
\ds \bar p \equiv p - \frac{1}{2\kappa} \left( V - F_{00} -
W\frac{a^2}{\bar a^2} \right) .
\end{array}
$$
Here, the $4$-velocity $u_{\mu}$ equals $\delta^0_{\mu}$ in
accordance with the assumed spatial isotropy, and $\bar a$ stands for
the scale factor of the metric $g_{\mu\nu}$. This is because the
solution for $g_{\mu\nu}$ is searched for in the form
$$
ds^2 = -dt^2 + \bar a^2(t)\left(dx^2+dy^2+dz^2 \right).
$$
(The scale factor $a(t)$ determines the vacuum metric
$g_{\mu\nu}^{(o)}$, as defined by (\ref{10}).) Now, the Einstein's
equations reduce to the familiar Friedman, and continuity equations
\begin{equation}\label{44}
\bar H = \frac{\kappa}{3}\,\bar\rho \,, \,\quad
\dot{\bar\rho} + 3\bar H \left(\bar\rho+\bar p\right) = 0 \,,
\end{equation}
where $\bar H \equiv \dot{\bar a}/\bar a$. The matter field equations
are represented by the equation of state
\begin{equation}\label{45}
p = w \rho \,.
\end{equation}
Now, one can use (\ref{44}) and (\ref{45}) to derive the differential
equation for the scale factor $\bar a$. As it turns out, it has the
form
\begin{equation}\label{46}
2{\bar a}\,\ddot{\bar a} + \left(3w+1\right) \dot{\bar a}^2 +
\alpha\,{\bar a}^2 + \left(3w+1\right) \beta = 0 \,,
\end{equation}
where the coefficients $\alpha$ and $\beta$ are defined by
$$
\alpha \equiv
-\frac{1}{2}\left[\left(w+1\right) V +
\left(w-1\right)F_{00}\right] ,  \quad
\beta \equiv \frac{1}{2}W a^2 .
$$
The fluid energy density $\rho$ becomes
\begin{equation}\label{47}
\rho = \frac{3}{2\kappa} \left[ 2\left(\bar H^2 - H^2\right) +
W a^2 \left( \frac{1}{\bar a^2} -
\frac{1}{a^2} \right) \right] .
\end{equation}

\begin{figure}[htb]
\begin{center}
\includegraphics[height=4cm]{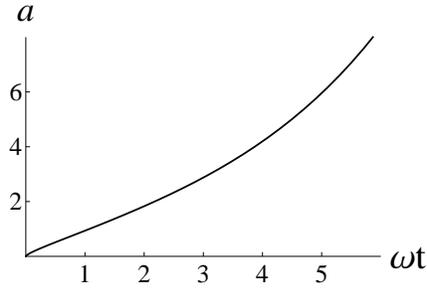}
\end{center}
\vspace*{-.5cm}
\caption{Scale factor modified by dust matter.\label{f4} }
\end{figure}

\begin{figure}[htb]
\begin{center}
\includegraphics[height=4cm]{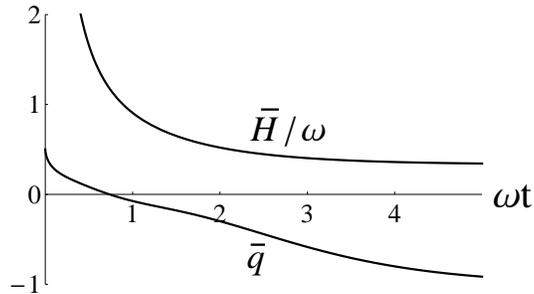}
\end{center}
\vspace*{-.5cm} \caption{Hubble and deceleration parameters in the
presence of dust matter.\label{f5} }
\end{figure}

The nonlinear differential equation (\ref{46}) is too complicated to
be solved analytically. For this reason, I solved it numerically for
a number of initial conditions. The simplest solution is obtained if
we choose $\bar a(0)=1$, $\,\dot{\bar a}(0)=0$, which yields
$$
{\bar a}(t) = a(t) \,, \quad \rho =0
$$
for every $w$. This simple solution can easily be verified
analytically. It states that the absence of ordinary matter
($\rho=0$) brings us back to the geometric vacuum (\ref{10}). The
nontrivial solutions are obtained for other choices of initial
conditions. But first, let me point out that a straightforward
analysis of (\ref{46}) shows that regular solutions with everywhere
positive $\rho$ do not exist. The regular solutions for $\bar a$ turn
out to oscillate around $a$, thereby causing $\rho$ to oscillate
around $\rho =0$. The situation is similar to that of the
$\Lambda$CDM model. To avoid negative $\rho$, one must reconcile with
the existence of singularities. As an example, I shall consider dust
matter ($w=0$), and the initial conditions $\bar a(0)\ll 1$,
$\,\dot{\bar a}(0)\gg\omega$. These initial conditions ensure that
singularity resides at $t\approx 0$, and determine the amount of
matter that agrees with observations. For $\bar a(0) = 10^{-9}$,
$\,\dot{\bar a}(0) = 10^4\,\omega$, the graph of the scale factor
$\bar a(t)$ is displayed in Fig. \ref{f4}.
%
%
The present time $t_0$, and the value of the parameter $\omega$ are
determined from the observed values of the Hubble and deceleration
parameters. The graphs of $\bar H(t)$ and $\bar q(t)$ are displayed
in Fig. \ref{f5}.
%
%
Unlike the matter-free values of $H(t)$ and $q(t)$, displayed in Fig.
\ref{f3}, the modified Hubble and deceleration parameters $\bar H(t)$
and $\bar q(t)$ have the solution $t=t_0$ for which $\bar q(t_0) =
-1/2$, $\,\bar H(t_0)>0$. One straightforwardly reads the values
$$
\omega t_0 = 2.70 \,,\quad \bar H_0 = 0.43\ \omega \,,
$$
which lead to
$$
\omega = 3.7 \cdot 10^{-33}\ {\rm eV} \,,\quad
t_0 = 15.4\ {\rm Gyr} \,.
$$
Now, it is straightforward to determine the graviton mass at the
present time. One finds
$$
m_g = 2.3 \cdot 10^{-33}\ {\rm eV} \,,
$$
which is only a fraction larger than its lower bound established in
Sec. \ref{Sub4a}. In the next section, I shall provide an example of
the Universe whose background value of $q_0$ is smaller than $-1$.
With the proper choice of the potential, the graviton mass will go
far below the bound of Sec. \ref{Sub4a}.

Finally, let me calculate the present value of the energy density
$\rho$, as given by the equation (\ref{47}).
%
\begin{figure}[htb]
\begin{center}
\includegraphics[height=4cm]{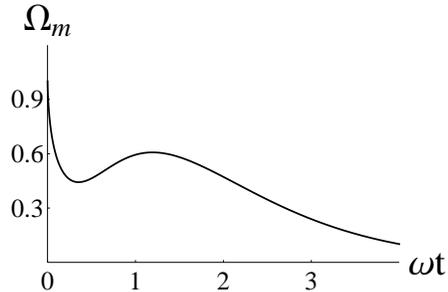}
\end{center}
\vspace*{-.5cm} \caption{Energy density of dust matter.\label{f6}}
\end{figure}
%
The corresponding density parameter $\Omega_m$ is defined by
$\Omega_m = {\rho}/{\rho_c}$, where $\rho_c \equiv {3\bar
H^2}/{\kappa}$ is the critical density at time $t$. The graph of the
function $\Omega_m(t)$ is shown in Fig. \ref{f6}. Given the present
time $\omega t_0 = 2.7$, the current value of $\Omega_m$ is
straightforwardly read to be
$$
\Omega_m(t_0) \approx 0.3 \,.
$$
One should have in mind, however, that the value of $\Omega_m$ is
very sensitive to the initial conditions chosen to solve the equation
(\ref{46}). The value $\Omega_m = 0.3$ is obtained with the choice
$\bar a(0) = 10^{-9}$, $\,\dot{\bar a}(0) = 10^4\,\omega$. For
different choices of initial conditions, one can obtain both, smaller
and larger values of $\Omega_m$.

\subsection{Bouncing Universe with negligible graviton mass}\label{Sub5c}

\begin{figure}[htb]
\begin{center}
\includegraphics[height=3.7cm]{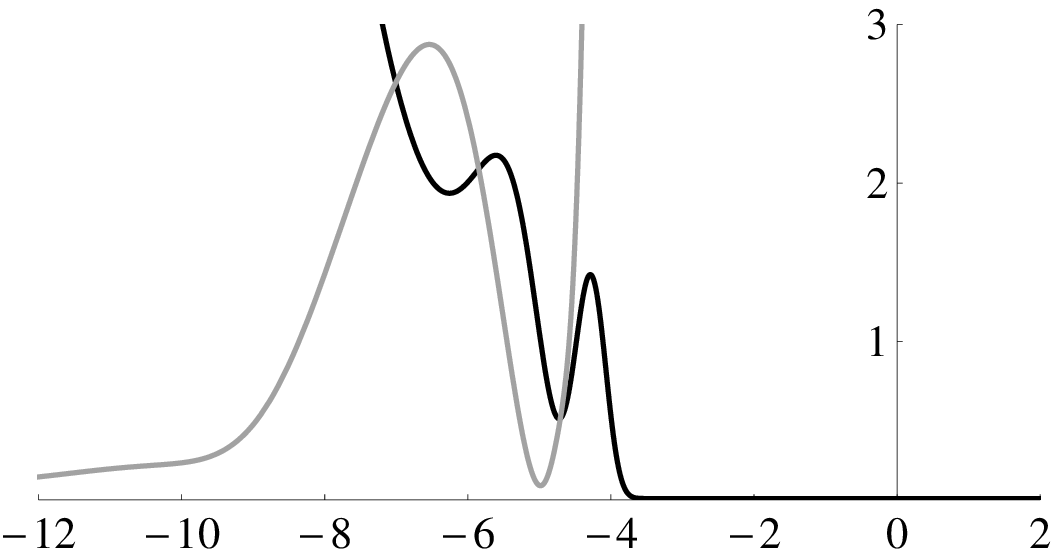}
\end{center}
\vspace*{-.5cm} \caption{Graphs of $\det{\cal M}_1$
(black), and $\left({\cal M}_1\right)_{11} +
\left({\cal M}_1\right)_{22}-
2\sqrt{\det{\cal M}_1}$ (gray). \label{f7}}
\end{figure}

In this subsection, I shall examine the bouncing Universe defined by
the scale factor
\begin{subequations}\label{48}
\begin{equation}\label{48a}
a={\ds e^{\,\omega t}}+\ln\left[1+{\ds e^{-\left(\omega t+9\right)}}\right],
\end{equation}
and the potential
\begin{equation}\label{48b}
W = -2 \omega^2 \exp\left[{\ds -e^{2 \left(\omega t+5\right)}}\right].
\end{equation}
\end{subequations}
It is straightforward to verify the regularity conditions (\ref{13}),
but it is not so easy to prove stability. Indeed, the potential
(\ref{48b}) is not of the general form (\ref{36}), for which the
general stability arguments of Sec. \ref{Sec4} hold. This leads us to
check the stability conditions (\ref{35}) by direct calculation.
First, I change the variables from $\Psi$ to $\Psi_1$, as defined in
(\ref{28.5}). The transformed mass matrix ${\cal M}_1$ is calculated
from (\ref{29.5}). Once the matrix elements of ${\cal M}_1$ are
known, one can determine the expressions $\det{\cal M}_1$ and
$\left({\cal M}_1\right)_{11} + \left({\cal M}_1\right)_{22} -
2\sqrt{\det{\cal M}_1}$ to see if they are everywhere non-negative,
as required by (\ref{35}). As opposed to the case $ W =
-\omega^2/a^2$, these expressions are very complicated, and
practically unusable for analytic studies. Instead, a computing
program is used for drawing their graphs, and establishing their
positivity. The graphs are displayed in Fig. \ref{f7}.
%
%
It is seen that the conditions (\ref{35}) are everywhere satisfied,
and therefore, the stability of the model is verified.

The model under consideration shares many features with the model of
the preceding subsection. For this reason, I shall restrain from
drawing graphs, and elaborating details. Let me just describe the
basic features of the model. For one thing, it is a bouncing Universe
with a linear contraction, and exponential expansion. The bounce is
located at $\omega t \approx -4.5$. Both, the Hubble parameter $H$,
and the deceleration parameter $q$ are monotonously increasing
functions after the bounce. In particular, $q(t)<-1$ during the whole
expanding phase. As in the preceding example, the inclusion of dust
matter increases $q(t)$ to take values well above $q=-1/2$. The
differential equation (\ref{46}) has been solved numerically. With
the initial conditions $\,\bar a(0) =10^{-3}$ and $\,\dot{\bar
a}(0)=10^3\,\omega$, one finds
$$
\omega = 1.3 \cdot 10^{-33}\ {\rm eV} \,,\quad
t_0 = 12.4 \ {\rm Gyr} \,.
$$
With these data, one can readily calculate the present energy density
$\rho_0$, and the present value of the graviton mass. The former is
found to satisfy $\kappa\rho_0 = 1.5\,\omega^2$, wherefrom one
finds
$$
\Omega_m \approx 0.3 \,.
$$
The latter is calculated by evaluating the potential $W$ in $t=t_0$.
Thus, one obtains
$$
m_g = 1.8 \cdot 10^{-22084}\ {\rm eV} \,.
$$
This result shows that graviton mass can indeed be arbitrarily small.
Precisely, the class of geometric sigma models considered in this
paper contains a subclass characterized by the arbitrarily small
graviton mass.

Before I close this subsection, let me comment on other masses that
appear in the particle spectrum. The vector modes, as governed by the
equation (\ref{22}), have the same mass as tensor modes. Thus, their
masses are as negligible as the graviton mass. However, their
friction term is significantly different. Indeed, when calculated in
$t=t_0$, it takes the value $2\cdot 10^5\, H_0$, which is five orders
of magnitude larger than that of the graviton. With such a big
friction term, vector particles have probably decayed long time ago.
As for the scalar modes, their mass squares are the eigenvalues
$\lambda_{\pm}$ of the mass matrix ${\cal M}_1$. They are defined by
$$
{\arraycolsep 0em
\begin{array}{rl}
\ds \lambda_{\pm} = \frac{1}{2} \bigg\{ & \ds \left({\cal M}_1
\right)_{11} + \left({\cal M}_1\right)_{22}                   \\
& \ds \pm \sqrt{\big[\left({\cal M}_1\right)_{11} +
\left({\cal M}_1\right)_{22}\big]^2 - 4\det{\cal M}_1} \bigg\} .
\end{array}}
$$
The present time values of the masses are found by calculating
$\lambda_{\pm}$ at $t=t_0$. One obtains
$$
m' \approx  10^{-30}\ {\rm eV} \,, \quad
m'' \approx 0.8 \cdot 10^{-22079}\ {\rm eV} \,.
$$
At the same time, the respective friction coefficients $f'$ and $f''$
are obtained by calculating the eigenvalues of the matrix ${\cal
F}_1$. They take the values
$$
f' \approx 2\cdot 10^5\, H_0 \,, \qquad
f'' \approx H_0 \,.
$$
It is seen that the heavier scalar mode quickly decays, and we are
left with one scalar mode of negligible mass. The effective particle
spectrum thus consists of one scalar and two tensor massless modes.

At the end, let me note that the above result is not unexpected.
Indeed, the potential $W$, as defined by (\ref{48b}), has an
extremely fast approach to zero as $t\to\infty$. The zero value of
$W$, on the other hand, makes the target metric $F_{ij}$ degenerate,
as seen from the definition (\ref{11}). The sigma model (\ref{8})
then reduces to the sigma model with only one scalar field. As a
consequence, the resulting dynamics at late times is expected to
carry three effective degrees of freedom. This is an example of
practical impossibility to distinguish between truly massless fields
and those with extremely small masses.

\section{Recapitulation}\label{Sec6}

I have shown in this paper that every bouncing metric can be a stable
solution of a simple geometric sigma model. In many respects, these
models look like ordinary sigma models that govern dynamics of four
scalar fields minimally coupled to gravity. What makes them different
is the way they are constructed. Namely, one first chooses the metric
one would like to be the vacuum of the model, and then builds up the
theory such that this metric becomes one of its solutions. The
procedure that associates an action functional with every
homogeneous, isotropic and spatially flat geometry is explained in
Sec. \ref{Sec2}. It is seen that all four scalar fields of the model
can be gauged away, leaving us with a purely metric theory. This is
the reason these models are called geometric.

The construction scheme of Sec. \ref{Sec2} does not guarantee that
small perturbations of the vacuum have stable dynamics. In fact, the
resulting indefiniteness of the sigma model target metric suggests
the opposite. This led me to perform a separate stability analysis.
First, in Sec. \ref{Sec3.5}, I have proven that singularities that
appear in the field equations are nonphysical. In particular, I have
shown that all perturbations formed in the past regularly pass the
bounce. In Sec. \ref{Sec4}, their stability is proven. Specifically,
for every background value of the scale factor $a(t)$ there is a
class of potentials $W(t)$ that makes the dynamics of small
perturbations stable. Vacuum stability against matter fluctuations is
only shortly mentioned, because it is technically identical to that
of Ref. \cite{33}. The important conclusion is that ordinary matter
does not compromise the stability of geometric sigma models.

In Sec. \ref{Sec5}, three simple examples are considered. The first
has only been used for the demonstration of how the procedure
described in Sec. \ref{Sec2} works in practice. The second example is
about slowly contracting Universe that has exponential expansion
after the bounce. The measured values of the Hubble and deceleration
parameters are obtained after ordinary dust matter has been included.
In this respect, this example resembles the $\Lambda$CDM model. The
graviton mass is calculated to be more than ten orders of magnitude
smaller than its upper bound reported by the LIGO experiment
\cite{45}. The third example has been included to show that, in some
cases, the graviton mass can be made arbitrarily small. In this
particular example, an extremely high friction coefficient makes the
two vector and one scalar mode decay quickly after the bounce. Only
two tensor modes, and one scalar mode survive. As it turns out, their
masses are of the order $10^{-22000}$ eV. Moreover, with an
appropriate choice of the potential $W$, one can make these masses
arbitrarily small.

Before I close this section, let me say something about physical
consequences of the considered multi-scalar cosmological models.
Specifically, a curious reader might be interested in what kind of
experiment is needed to justify (or rule out) the suggested class of
models. There are several observational possibilities to distinguish
my multi-scalar theory from single-scalar theories commonly discussed
in literature. First, as contrasted with the typical single-scalar
models, the graviton in my model is necessarily massive. As the
present time value of the graviton mass goes far beyond its
observational bound, the only way to measure modifications caused by
the graviton mass is the detection of possible anomalies in the
behavior of large cosmic structures. We already know that massive
graviton can only weaken the gravitational force at large distances,
which is quite the opposite of what one needs to explain the anomalous
galactic curves. This leaves us with clusters of galaxies, or voids,
as possible candidates for anomalous behavior due to graviton mass.
Second, as opposed to massive graviton, the scalars of the theory
typically enhance gravitational force. Thus, one may hope that a
properly defined cosmological scalar-tensor theory may explain the
anomalous galactic curves. Obviously, multi-scalar theory is expected
to be more effective than a single-scalar theory, when it comes to
modification of the gravitational force. Additional calculations along
these lines are needed for the full comparison of the models. The
third possibility is the comparison of calculated cosmological
parameters. The problem with this is that the considered class of
geometric sigma models has a subclass which is arbitrarily close to
the class of single-scalar models. As shown in the example
\ref{Sub5c}, the two vector modes, and one scalar mode of this
subclass have extremely high friction terms, so that they quickly
decay after the bounce. The remaining 2 tensor and 1 scalar mode have
negligible masses, which makes them practically indistinguishable from
typical single-scalar degrees of freedom. As a consequence, if cosmological
parameters of a single-scalar theory agree with observations, the
difference between the two models cannot be established by mere
comparison of calculated parameters. Otherwise, the observed
cosmological parameters might clearly distinguish between the two
theories. To examine this possibility, additional calculations are
needed. Finally, one can try to detect the predicted scalar and vector
particles in collisions of high energy particles. Owing to the extreme
weakness of their coupling to ordinary matter, this seems unlikely to
happen in the near future.

To summarize, I have shown in this paper that any bouncing metric can
be a stable solution of a simple model. One hopes that the class of
bouncing cosmologies thus obtained could accommodate a viable
cosmological model one searches for. The realization of this program
is expected from future investigations along these lines.

\begin{acknowledgments}
This work is supported by the Serbian Ministry of Education, Science
and Technological Development, under Contract No. $171031$.
\end{acknowledgments}

\end{document}